\documentclass[11pt,a4paper]{article}
\RequirePackage{amsmath,amssymb}
\RequirePackage[dvipsnames,usenames]{color}

\usepackage{cite}
\usepackage{fullpage}

\usepackage[british]{babel}
\usepackage[latin1]{inputenc}
\usepackage[T1]{fontenc}
\usepackage[final]{showkeys} 
\usepackage[bookmarks]{hyperref}
\usepackage{amsthm}
\usepackage{graphicx}
\usepackage{subfigure}
\usepackage{braket}
\usepackage{mathrsfs}

\usepackage{bm}
\usepackage{amsfonts}
\usepackage{braket}
\usepackage{graphics}

\newcommand{\bc}{\begin{center}}
\newcommand{\ec}{\end{center}}

\newcommand{\be}{\begin{equation}}
\newcommand{\ee}{\end{equation}}

\title{\bf Lagrangian formulation, a general relativity analogue, and a 
symmetry of the Vialov equation of glaciology}

\author{
Valerio~Faraoni 
$\,$
\\
\\
Department of Physics \& Astronomy, Bishop's University
\\
2600 College Street, Sherbrooke Qu\'ebec, Canada J1M~1Z7
}

\begin{document}
\maketitle
\begin{abstract}

Using a suitable rescaling of the independent variable, a Lagrangian is 
found for the nonlinear Vialov equation ruling the longitudinal profiles 
of glaciers and ice caps in the shallow ice approximation. This leads to a 
formal analogy between the (rescaled) Vialov equation and the Friedmann 
equation of relativistic cosmology, which is explored. This context 
provides a new symmetry of the (rescaled) Vialov equation and gives, at 
least formally, all its solutions using a generating function, which is 
the Nye profile for the degenerate case of perfectly plastic ice. 
\end{abstract}

\newpage

\section{Introduction}
\label{sec:1}
\setcounter{equation}{0}

An important geomorphological feature of ice caps and alpine glaciers is 
their longitudinal profile, {\em i.e.}, the ice thickness $h$ as a 
function of a coordinate $x$ running downstream along the glacier bed. 
Some knowledge of longitudinal profiles is needed in 
estimating the volume of a glacier or ice sheet \cite{FaraoniJOG16}, 
and the volume-area scaling for these systems constitutes a major tool in 
estimating ice content, ice loss, and sea level rise related to climate 
change \cite{Bahrreview}.

The mathematical modelling of longitudinal glacier profiles is relatively 
well developed and is based on non-linear 
differential equations.  The rheology 
of glacier ice plays a crucial role in determining the shape $h(x)$ of 
an ice cap or valley glacier. The response of glacier ice 
to applied stresses is well described by Glen's law relating the strain 
rate tensor  $\dot{\epsilon}_{ij}$ with the stresses in the ice 
\cite{Glen55}
\be\label{Glen}
{\dot{\epsilon}}_{ij} = {\cal A} \, \sigma_\text{eff}^{n-1} s_{ij}
\ee
where ${\cal A}$ is a (temperature-dependent) constant 
\cite{Paterson94,CuffeyPaterson10,Hooke05,GreveBlatter09}, 
$\hat{s} = \left( s_{ij} \right) $ is the deviatoric stress tensor, and 
\be
\sigma_\text{eff} =\sqrt{ \frac{1}{2} \mbox{Tr}\left( 
\hat{s}^2 \right)} 
\ee  
is the effective stress.  Glen's 
law describes the rheology of ice.  The strain rate 
$\dot{\epsilon}_{ij}$  
depends on the deviatoric stresses in all directions through 
$\sigma_\text{eff}$. Different values of the exponent $n$
corresponds to very different physical situations. At very low stresses,
the value $n=1$ corresponds to a perfectly viscous material with viscosity
coefficient $\eta= {\cal A}^{-1}$ and satisfying the linear relation $
s_{ij}=\dot{\epsilon}_{ij}/{\cal A}$. The value $n=2$ is sometimes adopted
to
describe basal sliding of a glacier. The value $n=3$ corresponds to ice
flow, and is the one commonly adopted in the modelling of ice flow in  
glaciers.  The limit $n\rightarrow +\infty $ corresponds to perfectly  
plastic ice, a poor approximation to real ice flow
\cite{Paterson94,CuffeyPaterson10,Hooke05,GreveBlatter09}.

Longitudinal profiles $h(x)$ are modelled under the assumptions of  
incompressible and  isotropic ice, steady state, a flat bed 
(a plane which, in general, has
 non-zero slope), the shallow ice approximation, and Glen's law to 
describe the ice behaviour 
\cite{Paterson94,Hooke05,CuffeyPaterson10,GreveBlatter09} (these are major 
simplifications and modern numerical models attempt to go beyond these 
approximations). Then, 
the longitudinal glacier profile $h(x)$ 
satisfies the Vialov ordinary differential equation 
\cite{Vialov58,Paterson94,CuffeyPaterson10,Hooke05,GreveBlatter09}
\be\label{Vialov}
x \, c(x) = \frac{2{\cal A} }{n+2} \left( \rho_\text{ice} g h \left| 
\frac{dh}{dx} \right| \right)^{n} h^2 \,,
\ee
where $\rho_\text{ice}$ is the ice density, $g$ is the acceleration of 
gravity, $c(x)$ is the accumulation rate of ice, {\em i.e.}, 
the flux density of ice volume in the $z$-direction, which is  
perpendicular to the $x-$axis and the flat bed ($c$ has the 
dimensions of a velocity). The Vialov equation is essentially the mass  
conservation equation in the shallow ice approximation for flat glacier  
bed.

Analytic expressions describing longitudinal glacier 
profiles are needed in studying several aspects of glaciology 
({\em e.g.}, 
\cite{Thorp91,Ngetal10,BennHulton10,Bahrreview,FaraoniJOG16}). 
It is customary 
to search for solutions in the finite interval 
$x \in 
\left[0,L \right] $, where the glacier summit is located at $x=0$ and 
the terminus is at $x=L $, and $L$ is the length of the glacier or ice 
cap. In this case $dh/dx$ is negative and its absolute value must be 
taken in the Vialov equation. One could switch the 
location of summit and terminus, then $dh/dx>0$ in $\left(0, L \right)$ 
and the  absolute value appearing in Eq.~(\ref{Vialov}) takes care of both 
situations. 

For ice caps and ice sheets, once a solution for the longitudinal profile 
of half of a glacier is found in $\left[ 0, L \right]$, it is extended by 
continuity to the interval $\left[ -L, L \right]$ (or to $\left[0, 2L 
\right]$, respectively) by reflecting it about the vertical line $x=0$ (or 
$x=L$, respectively) passing through the summit, where the diffusion 
degenerates.  The surface profile $h(x)$ of an ice cap or ice sheet is 
continuous but not differentiable at the summit, since the left and right 
derivatives of $h$ have opposite signs there.

At the terminus, the slope $dh/dx$ and the basal stress $\tau_b= 
-\rho_\text{ice} g h \, dh/dx$ diverge; this well known shortcoming of the 
modelling is due to the breakdown of the shallow ice approximation at the 
terminus \cite{Paterson94,CuffeyPaterson10,Hooke05,GreveBlatter09} (this 
is not  a general feature of the shallow ice approximation).

The formal solution of the Vialov equation~(\ref{Vialov}) is 
\be \label{h(x)}
h(x)=\left\{ \mp \frac{ 2\left(n+1\right)}{ n\rho_\text{ice} g} \left( 
\frac{n+2}{2{\cal A}} \right)^{1/n} 
\int dx \left[ x \, c(x) \right]^{1/n} 
\right\}^{\frac{n}{2(n+1)} } \equiv A  
\left[ I(x) \right]^{\frac{n}{2(n+1)} } \,, 
\ee
where the upper sign applies to the case in which the summit is at $x=0$ 
(and  $dh/dx<0$), and the lower sign  if $x=L$ is the 
summit. Here
\be
A  \equiv  \left[ \frac{ 2\left(n+1\right)}{n\rho_\text{ice} g} 
\left(\frac{n+2}{2{\cal A}} \right)^{1/n}\right]^{\frac{n}{2(n+1)}}  \,,
\ee
and the integral
\be
I(x)  \equiv \int dx \left[ x \, c(x) \right]^{1/n} \,.
\label{I}
\ee
is defined up to an arbitrary integration constant. 
A function $c(x)$ modelling the accumulation rate of ice 
must be prescribed in the model, ideally based on atmospheric models for 
precipitation in the region. Even simple choices of $c(x)$ make the 
integral~(\ref{I})  impossible to compute explicitly in terms of 
elementary 
functions. A few analytic 
solutions of the Vialov equation are available 
\cite{Bodhvardsson55,Vialov58,Weertman61,Paterson72,Bueler03,Bueleretal05}.
The Chebysev theorem of integration can be 
used to characterize the models in which the 
solution can be expressed in terms of elementary functions \cite{myMG17}.  
Other analytic profiles  follow from the rather 
unrealistic assumption of perfectly plastic ice used in the 
early modelling, and only appropriate when the 
deformation of the ice is negligible  
\cite{Nye51a,Nye51b,FaraoniVokey15,FaraoniEJP19}, which is formally 
obtained as the limit $n\rightarrow +\infty$ of the Vialov equation 
\cite{Paterson94,CuffeyPaterson10,Hooke05,GreveBlatter09}.

As shown in the next section, by suitably rescaling the independent 
variable $x$, it is found that the Vialov equation admits Lagrangian and 
Hamiltonian formulations, thus solving an inverse variational problem. 
This is rather surprising because the Vialov equation was not derived 
originally from a variational principle or from the extremization of a 
physical quantity. Finding a Lagrangian and a Hamiltonian from a 
given equation constitutes the inverse variational problem of 
mathematical physics (sometimes called Helmoltz problem \cite{Helmoltz}), 
which can be solved in a surprisingly wide number 
of cases \cite{Lopuszanski} and, recently, has been approached using 
non-standard Lagrangians  for dissipative-like autonomous 
differential equations ({\em e.g.}, 
\cite{MusielakRoySwift08,Musielak08,CieslinskiNikiciuk10,SahaTalukdar14}). 
Similar to some of the equations considered in 
Ref.~\cite{SahaTalukdar14}, our procedure is based on a redefinition of 
variables.

Once a Hamiltonian is available, one quicky realizes 
that it is conserved and the energy conservation equation leads to a 
straightforward analogy between the profile $h(x)$ solving the (rescaled) 
Vialov equation and the motion of a particle in one dimension under a 
suitable potential energy. What is more, this formulation opens the door 
to a formal analogy between the Vialov equation (using the rescaled bed 
coordinate) and the Friedmann equation of spatially homogeneous and 
isotropic cosmology. This analogy is explored in Sec.~\ref{sec:4} and used 
in Sec.~\ref{sec:5} to generate all the solutions of the Vialov equation, 
at least formally, from a generating function which is the Nye parabolic 
profile for perfectly plastic ice.  Section~\ref{sec:6} contains the 
conclusions.

\section{Lagrangian formulation of the Vialov equation}
\label{sec:2}
\setcounter{equation}{0}

Rewrite the Vialov equation~(\ref{Vialov}) as 
\be
h^{ \frac{n+2}{n}} h'= \left[ \frac{ x \, c(x)}{ \alpha}\right]^{1/n} 
\,,
\ee
where 
\be
 \alpha = \frac{2{\cal A}}{n+2} \, \left( \rho_\text{ice} g \right)^n \,.
\ee 
Now change the independent variable, $x \rightarrow \bar{x}$,  
with $\bar{x}$ defined  by 
\be
d\bar{x}= \left[ x \, c(x) \right]^{1/n} dx \,, \label{eq:barx}
\ee
or
\be
\bar{x} (x) =\int_0^x dx' \, \left[ x' c(x') \right]^{1/n} \,.
\ee
This relation is invertible and one-to-one  
where $c(x)>0$, then $d\bar{x}/dx>0$; $c(x)$ is 
required to be continuous and piecewise differentiable.\footnote{The 
exception  in the literature is the Weertman-Paterson model 
\cite{Weertman61,Paterson72} 
in which $n=3$ and $c(x)$ 
is a step 
function defined on two intervals $\left(0, R \right)$ and $\left( R,L 
\right)$ on which it is piecewise 
constant and has opposite sign. In this case one studies the two 
intervals separately and reduces to the previous situation.}

With the new independent variable $\bar{x}$, the Vialov equation assumes 
the form
\be
h^{\frac{n+2}{n}} \left| \frac{dh}{d\bar{x}} \right| = 
\frac{1}{\alpha^{1/n}} \,.\label{assumes}
\ee
Dividing both sides by $ \sqrt{2} \, h^{\frac{n+2}{n}}$ and squaring, one 
obtains
\be
\frac{1}{2} \, \left( \frac{dh}{d\bar{x}}\right)^2 = 
\frac{1}{2\alpha^{2/n}} \, h^{-\frac{2(n+2)}{n} } \,,
\ee
which looks like an energy conservation equation for a particle of unit 
mass and position $h$ in one-dimensional motion in the potential energy
\be
V( h)= - \frac{V_0}{h^{ \frac{2(n+2)}{n} } } \,, \;\;\;\;\;\;\;\;
V_0= \frac{1}{2 \alpha^{2/n} } \,,\label{potential}
\ee
with $\bar{x}$ playing the role of time. 
It is then 
intuitive to write the Lagrangian
\be
{\cal L}\left( h(\bar{x}), \frac{dh}{d\bar{x}} (\bar{x}) \right) =  
\frac{1}{2} 
\, \left( \frac{dh}{d\bar{x}} \right)^2 - 
\frac{V_0}{ h^{\frac{2(n+2)}{n} } }\,. \label{eq:Lagrangian}
\ee
In the limit of perfectly plastic ice $n\rightarrow +\infty$, the variables  
$\bar{x}$ and $x$ coincide, the accumulation function $c(x)$ disappears 
from the picture, and the Lagrangian~(\ref{eq:Lagrangian}) reduces to 
\be
{\cal L}_{\infty}\left( h(\bar{x}), \frac{dh}{d\bar{x}} (\bar{x}) \right) 
= \frac{1}{2}  \, \left( \frac{dh}{d\bar{x}}\right)^2 - 
\frac{V_0}{h^2} \,. 
\ee
In the general case with finite $n$, the Lagrangian~(\ref{eq:Lagrangian}) 
does not depend on $\bar{x}$ and the corresponding Hamiltonian is 
conserved, yielding the Beltrami identity 
\be
{\cal H}= \frac{\partial {\cal L}}{ \partial \left( dh/d\bar{x} \right)} 
\, 
\frac{dh}{d\bar{x}} - {\cal L} =  \frac{1}{2} \, \left( 
\frac{dh}{d\bar{x}}\right)^2 -\frac{1}{2\alpha^{2/n} \, 
h^{\frac{2(n+2)}{n} } } = E  \,, \label{eq:energy}
\ee
where $E$ is the constant energy of the analogous particle.  The Vialov 
equation is reproduced for vanishing energy $E$. 

When a solution $h(\bar{x})$ of this equation is found, one still has to 
replace $\bar{x}$ with $x$, which may not be possible to do explicitly 
using only elementary functions. The situations when the integral $I(x)$ 
can be reduced to elementary functions by means of the Chebysev theorem 
are discussed in \cite{myMG17}.

As an example, in the Vialov model  \cite{Vialov58} it is $c(x)=c_0 
\equiv$~const. and 
\be
\bar{x}(x) = \frac{n \, c_0^{1/n}  }{n+1} \, x^{ \frac{n+1}{n} } \,. 
\label{Vialovbarx}
\ee
Searching, as customary in the literature, for a solution in the interval 
$x\in \left[0,L\right]$ with the glacier summit at $x=\bar{x}=0$ (where 
$h(0)=H$)  and terminus at $x=L$ (where $h=0$ and, necessarily, 
$dh/dx \rightarrow \infty$), one has
\be
h^{ \frac{n+2}{n}} \, \frac{d h}{d\bar{x}} =-\frac{1}{\alpha^{1/n} } \,,\label{eq:V18}
\ee 
which integrates to 
\be
\left[ h(\bar{x}) \right]^{ \frac{2(n+1)}{n} } = 
H^{ \frac{2(n+1)}{n} } \left[ 1-\frac{2(n+1) \bar{x} }{ n \alpha^{1/n} 
H^{ \frac{2(n+1)}{n} }   } \right] \,.
\ee
Raising both sides to the power $\frac{n}{ 2(n+1) } $, one obtains the 
solution
\be
h(\bar{x} ) = H \left[ 1-\frac{ 2(n+1) \bar{x} }{ 
n \, \alpha^{1/n} H^{ \frac{2(n+1)}{n}}} \right]^{ \frac{n}{2(n+1)} } 
\ee
and finally, using Eq.~(\ref{Vialovbarx}) to return to the original 
independent variable $x$,
\be
h(x) = H \left[ 1-\left( \frac{x}{L}\right)^{\frac{n+1}{n}}  
\right]^{\frac{n}{2(n+1)} } \,,
\ee
where 
\be
L= \frac{1}{2^{ \frac{n}{n+1} } } \,   H^2 \left( \frac{\alpha}{c_0} 
\right)^{ \frac{1}{n+1} } \,,
\ee
which is the well known Vialov profile \cite{Vialov58}.

\section{Mechanical analogy}
\label{sec:3}
\setcounter{equation}{0}

\begin{figure*}
\center
  \includegraphics[width=0.40\textwidth]{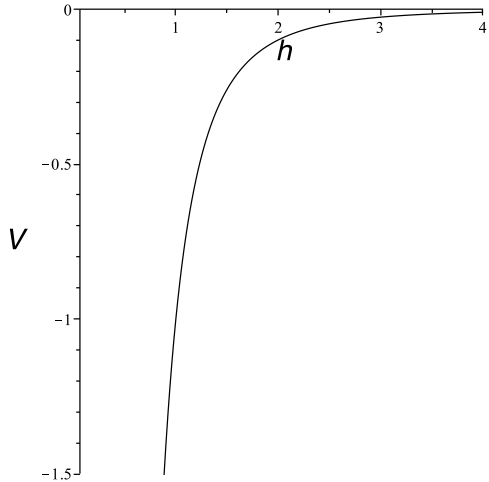}
\caption{The potential~(\ref{potential}) (plotted for $2\alpha^{2/n}=1$) 
for 
the realistic value $n=3$. Since the total energy is $E=0$, the particle 
is forced to remain on the $h$-axis.} \label{fig:V}      
\end{figure*}

Equation~(\ref{eq:energy}) has the form of an energy conservation equation 
for a particle of unit mass and total mechanical energy $E$ in the 
potential energy $ V(h)$ given by Eq.~(\ref{potential}). The 
Vialov equation is reproduced for the value $E=0$ of this energy. This 
brings about an obvious analogy. Since $n>0$ for all practical 
applications (and $n=3$ for ice creep), the potential $V(h)$ has the shape 
illustrated in Fig.~\ref{fig:V}, to which we refer.  

\begin{figure*} \center
  \includegraphics[width=0.40\textwidth]{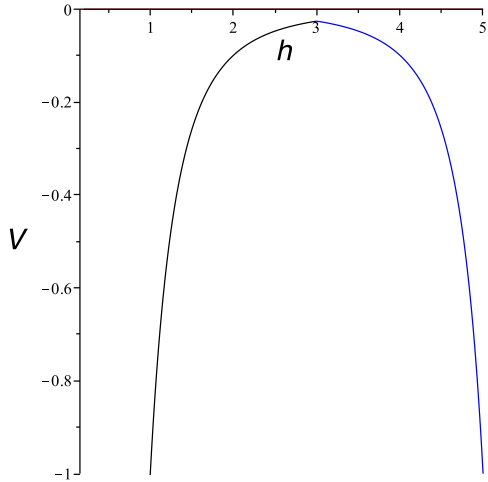}
\caption{The effective potential $V(h)$ (in black) is truncated at $h=H$ 
(arbitrarily set to $3$ here) and reflected about the vertical line $h=H$ 
(blue portion of the curve). The 
resulting potential and the solution $h(x)$ of the Vialov equation 
are not differentiable at $h=H$.} \label{fig:2}     
\end{figure*}

Only the interval 
$h \in \left[ 0, H \right]$ is relevant, where $V(h) \rightarrow -\infty$ 
as 
$h\rightarrow 0^{+}$ and $V(h) \leq V(H)<0$. The motion of the particle 
with total mechanical  energy $E=0$ is truncated artificially at the 
glacier summit where $h=H$ and the solution is 
extended by reflecting about 
the 
vertical line through the summit. In the $\left(h, V\right) $ plane, this 
corresponds to 
replacing $V(h)$ in $h\in(H, 2H) $ with the reflection of $V$ in 
$\left(0,H \right) $ about the line $h=H$ (see Fig.~\ref{fig:2}). The 
region $h> 2H$ is irrelevant.

Although, without this reflection, the motion of the particle would be 
unbounded, the truncation and the artificial confinement to $h\in \left(
0, L \right)$ change the picture from what would be deduced by applying 
the usual Weierstrass method.

One feature that emerges is that, in order for the particle to have finite 
total energy $E=0$, its kinetic energy must diverge as it falls into the 
origin $h=0$ corresponding to the glacier terminus, in order to compensate 
for the infinitely negative potential energy at $h=0$. This means that the 
derivative $dh/dx $ necessarily diverges at the terminus, a known feature 
of the solutions of the Vialov equation, which is associated with 
unphysically divergent stress at the bed $\tau_b= -\rho_\text{ice} g h \, 
dh/dx$. The shallow ice approximation, which is used to derive the Vialov 
equation, breaks down at the glacier terminus.

\section{Cosmological analogy}
\label{sec:4}
\setcounter{equation}{0}

Using the rescaled variable $\bar{x}$, the Vialov equation is formally 
analogous to the Friedmann equation describing spatially homogeneous and 
isotropic (or Friedmann-Lema\^itre-Robertson-Walker, hereafter FLRW) 
universes. Before unveiling the analogy, let us recall the basics of FLRW 
cosmology.

In general relativity, gravity is described by geometry, {\em 
i.e.}, by the Lorentzian metric tensor $g_{ab}$ which satisfies the 
Einstein 
field equations
\be
R_{ab}-\frac{1}{2} \, g_{ab} R=8\pi G T_{ab} \,, \label{efe}
\ee
where $R_{ab}$ is the Ricci tensor of the metric $g_{ab}$, $R$ is its  
trace, and the stress-energy tensor $T_{ab}$ describes the matter content 
of spacetime (we adopt the notation of Refs.~\cite{Wald84,Carroll02} using  
units in which the speed 
of light is unity, while $G$ denotes Newton's constant).
Spatially homogeneous and  isotropic 
universes have four-dimensional spacetime geometries described by the FLRW 
line element
\be
ds^2 = g_{ab} dx^a dx^b= -dt^2 +a^2(t) \left[ \frac{dr^2}{1-Kr^2} +r^2 
\left( d\theta^2 + 
\sin^2 \theta \, d\varphi^2 \right)\right] \,. \label{eq:10}
\ee
in comoving polar coordinates $\left(t, r, \theta,  \varphi 
\right)$. The 
scale factor $a(t)$ describes the expansion history of the universe. 
The constant $K$ is associated with  the constant 
curvature of the 3-dimensional Riemannian spatial geometries (the physical 
3-dimensional spaces) obtained by setting  
$dt=0$. If $K>0$, the FLRW line 
element~(\ref{eq:10}) belongs to a closed universe; if $K=0$, the  
spatial 3-sections are flat (Euclidean); if $K<0$, 3-space is  
hyperbolic  
\cite{Wald84,Carroll02,Liddle03,KT90}. The requirements of spatial 
homogeneity and isotropy are very stringent and, as a consequence, the 
previous classification includes all the possible FLRW geometries.
 
In FLRW cosmology, the matter content of the universe is usually described 
by a perfect fluid with stress-energy tensor of the form
\be
T_{ab}=\left( P+\rho\right) u_a u_b +P g_{ab} \,,
\ee
where $u_a$ is the fluid 4-velocity, $\rho(t)$ is the energy  density, and 
$P(t)$ is the isotropic pressure. Comoving coordinates are adapted to 
observers comoving with the fluid, who see the universe spatially 
homogeneous and isotropic around them (that is, comoving coordinates are 
adapted to the symmetries of the metric). $P$ and $\rho$ are related 
by an equation of state, usually of the barotropic form $P=P(\rho)$. Very 
often in the literature, this takes the form
\be
P=w\rho \,, \;\;\;\;\;\;\;\; w=\mbox{const.} \label{eqstate}
\ee
with constant ``equation of state parameter'' $w$.

The Einstein equations~(\ref{efe})  simplify greatly by requiring  spatial 
homogeneity 
and isotropy, and reduce to the Einstein-Friedmann equations
satisfied by $a(t), \rho(t)$, and $P(t)$ \cite{Wald84,Carroll02,Liddle03,KT90}  
\begin{eqnarray}
&&H^2 \equiv \left( \frac{\dot{a}}{a}\right)^2 =\frac{8\pi G}{3} \, \rho 
-\frac{K}{a^2} \,, \label{eq:11}\\
&&\nonumber\\
&&\frac{\ddot{a}}{a}= -\, \frac{4\pi G}{3} \left( \rho +3P \right) \,, 
\label{eq:12} \\
&&\nonumber\\
&& \dot{\rho}+3H\left(P+\rho \right)=0 \,.\label{eq:13}
\end{eqnarray}
Here an overdot denotes differentiation with respect to the comoving 
time $t$, while $H(t)\equiv 
\dot{a}/a$ is the Hubble function \cite{Wald84,Carroll02,Liddle03,KT90}.  
Among the  three equations (\ref{eq:11})-(\ref{eq:13}), only two are 
independent. Given any two of them, the third one is derived from them. 
Without loss of  generality, here we choose the Friedmann 
equation~(\ref{eq:11}) and the energy conservation equation~(\ref{eq:13}) 
as our independent equations, then the acceleration 
equation~(\ref{eq:12}) is derived from them. 
If the cosmic fluid obeys the equation of state~(\ref{eqstate}) with 
$w\neq -1$, the 
covariant conservation equation~(\ref{eq:13}) is integrated, 
giving
\be
\rho(a)=\frac{\rho_0}{a^{3(w+1)} }
\ee
and the Friedmann equation becomes
\be
H^2=\frac{8\pi G \rho_0}{a^{3(w+1)} } -\frac{K}{a^2} \,.
\ee

Going back now to the Vialov equation, Eq.~(\ref{eq:energy}) can be 
rewritten as 
\be
\left( \frac{1}{h} \, \frac{ dh}{d\bar{x}} \right)^2= 
\frac{1}{2\alpha^{2/n} \, h^{\frac{4(n+1)}{n} } } +\frac{E}{h^2} \,.
\ee
This is formally a Friedmann equation for a universe with scale factor 
$a(t)$ analogous to $h(\bar{x})$, curvature index 
$K=-E$, filled with a perfect fluid with constant equation of state 
parameter 
\be
w=\frac{n+4}{3n} \label{w}
\ee
and energy density $\rho(a) =\rho_0/a^{3(w+1)}=\rho_0 / a^{4(n+1)/n}$, 
with
\be
\rho_0 = \frac{3}{16\pi G \alpha^{2/n}} \,.
\ee
It is meaningful that the analogous energy density comes  out positive: 
this property is not to be taken for granted when building such analogies 
and its failure would take much value away from the analogy.

Ice creep,  corresponding to the value $n=3$ \cite{Glen55}, gives $w=7/9 
\simeq 0.778$. 
The equation of state parameter~(\ref{w}) is a positive and 
always decreasing function of $n$; it  
diverges at $n=0$ and has a horizontal asymptote  $w=1/3$ corresponding to 
the radiation era of the spatially flat universe and, in the analogy, to 
the limit of perfectly plastic ice $n\rightarrow +\infty$.

An inspection of the Friedmann equation shows that the Hubble function 
$H$ 
diverges when $a\rightarrow 0$, which signals a Big Bang (if the 
universe is expanding, $\dot{a}>0$) or a Big Crunch (if the universe is 
contracting, $\dot{a}<0$)  spacetime singularity. According to the 
Hawking-Penrose singularity theorems \cite{Wald84}, this singularity is 
generic and 
unavoidable if the cosmic fluid satisfies the weak energy condition 
$\rho>0$ and $\rho+3P>0$ (corresponding to $w>-1/3$ for a barotropic 
perfect fluid), which is always satisfied in our case. 

In the Vialov equation, the singularity in $h'$ at the terminus $x=L$ is 
due to the breakdown of the shallow ice approximation. In the cosmic 
analogue, the shallow ice approximation would mean small $\dot{a}$ or 
$a(t)/t \ll 1$. Since $a(t)=a_0 t^{ \frac{2}{3(w+1)} } $, 
we have 
that $ a(t)/t=a_0/ t^{ \frac{3w+1}{3(w+1)}} \rightarrow +\infty$ at the 
Big Bang or Big Crunch. While some cosmologists tend to think that quantum 
mechanics will change the behaviour of matter and circumvent the 
singularity theorems, thus avoiding the singularity, the analogy with the 
Vialov equation would
 suggest instead that the Friedmann equation only holds when the scale 
factor $a(t)$ is much smaller than the age of the universe $t$ at that 
time, $a(t) \ll t$, therefore away from the beginning. This intuition 
matches the idea that there is a fundamental length, the Planck length 
$\ell_{Pl}$ below which a continuous spacetime manifold does not exist.

Recent work has established that all the solutions of the Friedmann 
equation are roulettes \cite{Chen15b}. A roulette is the curve described 
by a point attached to a closed convex curve as that curve 
rolls without slipping along a second, given, curve.   This result, 
translated to the glaciology side of 
the analogy, states that all solutions of the rescaled Vialov equation are 
roulettes, a property hitherto undiscovered for this equation.

The Einstein-Friedmann equations for a spatially flat ($K=0$) universe and 
a perfect fluid with constant barotropic equation of state $P=w\rho$ enjoy 
the symmetry \cite{Symmetry2020}
\begin{eqnarray}
a &\rightarrow &\tilde{a}=a^s \,,\\
dt &\rightarrow & d\tilde{t}= s a^{ \frac{3(w+1)(s-1)}{2}} dt 
\,,\\
\rho &\rightarrow & \tilde{\rho}= a^{-3(w+1)(s-1)} \rho \,,
\end{eqnarray}
where $s\neq 0 $ is a real number. These symmetry transformations form 
a one-parameter Abelian group as $s$ varies in $\mathbb{R} 
\mathbin{\vcenter{\hbox{$\scriptscriptstyle\setminus$}}} \left\{0\right\}$ 
\cite{Symmetry2020}. Each symmetry transformation translates into the 
symmetry of 
the Vialov equation~(\ref{eq:V18}) written using the variable $\bar{x}$
\begin{eqnarray}
h &\rightarrow &\tilde{h}=h^s \,,\label{symmV1}\\
d\bar{x} &\rightarrow & d\tilde{x}= s \, h^{ \frac{2(n+1)(s-1)}{n}} 
d\bar{x} \,.\label{symmV2}
\end{eqnarray}
It is straightforward to check that the left hand side of 
Eq.~(\ref{eq:V18}) is invariant under the transformations~(\ref{symmV1}) 
and (\ref{symmV2}), while the right hand side is invariant because it is 
constant. Again, the transformations form a one-parameter Abelian group as 
$s\neq 0$ 
varies.

In the limit of perfectly plastic ice $ n \rightarrow +\infty$, the Vialov 
equation (written using the original variable $x$) can be written as 
\be
h |h'| h^{2/n} = \frac{ (n+2)^{1/n} \left( x c(x) \right)^{1/n} }{\left( 
2{\cal A} \right)^{1/n} \rho_\text{ice} g} \rightarrow \frac{1}{\rho_\text{ice} g} 
\ee
using $ \lim_{n\rightarrow +\infty} n^{1/n}=1$. The actual equation for 
perfectly plastic ice is 
\be
h |h'|= \frac{\tau_b}{\rho_\text{ice} g} \,, \label{Nyeequation}
\ee
where the constant $\tau_b$ is the stress at the bottom of the ice, and 
can be obtained by using Eq.~(\ref{eq:energy}) with the Lagrangian 
$ {\cal L}_{\infty}$. The limit $n\rightarrow +\infty$ of the symmetry 
produces 
\begin{eqnarray}
h &\rightarrow &\tilde{h}=h^s \,,\\
dx &\rightarrow & d\tilde{x}= s \, h^{ 2(s-1) } dx \,,
\end{eqnarray}
and it is straightforward to check, again, that it leaves invariant the 
left hand 
side of Eq.~(\ref{Nyeequation}), while the right hand side is invariant 
because it is constant. The solution of Eq.~(\ref{Nyeequation}) is the 
parabolic Nye profile \cite{Nye51a}
\begin{eqnarray}
h(x)&=& H\sqrt{ 1-\frac{x}{L}} \,,\\
&&\nonumber\\
H&=& \sqrt{ \frac{2\tau_b L}{\rho_\text{ice} g} } \,.
\end{eqnarray}
It is well known that the parabola is a roulette described by a point 
fixed on the involute of the circle rolling along a straight line.

\section{Reducing the Vialov equation to the Nye equation}
\label{sec:5}
\setcounter{equation}{0}

The symmetry of the Friedmann equation for $K=0$ with a perfect fluid with 
constant barotropic equation of state essentially says that one can 
obtain the solution for a fluid from that for another fluid. On the other 
side of the analogy, this property  is useful to approach the 
problem of solving  
the Vialov equation for general $n$. Noting that the equation for 
perfectly plastic ice ($n\rightarrow+\infty$) is simpler than the Vialov 
equation for finite $n$, we can try to reduce the problem of solving
the latter to that of solving the former. Indeed, the Vialov equation can 
be reduced to the Nye equation~(\ref{Nyeequation}) 
for perfectly  plastic ice by making one more change of variable. 
Keeping $\bar{x}$ given by Eq.~(\ref{eq:barx}) as the independent 
variable, let us change also the dependent variable according to $h 
\rightarrow \bar{h}=h^p$. Then the 
Vialov equation~(\ref{assumes}) becomes
\be
\bar{h}^{ \frac{ 2(n+1)-np}{np} } \, \frac{d\bar{h}}{d\bar{x}} 
=\frac{p}{\alpha^{1/n}} \,.
\ee
The choice 
\be
p=\frac{n+1}{n}
\ee
transforms the Vialov equation into the Nye equation
\be
\bar{h} \left| \frac{d\bar{h}}{d\bar{x}} \right|= 
\frac{1}{\bar{\alpha}^{1/n}} 
\ee
for $\bar{h}( \bar{x})$, where $\bar{\alpha}= \left(\frac{n}{n+1}\right)^n 
\alpha$. This property is in principle useful to generate solutions of the 
Vialov equation. For example, in the Vialov model with $c(x)=c_0=$~const., 
we have already seen that $\bar{x}=\frac{n c_0^{1/n}}{n+1} \, 
x^{\frac{n+1}{n}} $. Defining now $\bar{h}=h^{ \frac{n+1}{n}}$, the Vialov 
equation reduces to the Nye equation, which has as solution the Nye 
profile
\be
\bar{h}(\bar{x}) = \bar{H} \sqrt{ 1-\frac{\bar{x}}{\bar{L}} } 
\,.\label{barredNye}
\ee
This solution can be now be written in terms of $h$ and $x$ as
\be
h^{\frac{n+1}{n} } = \bar{H} \sqrt{ 1-\frac{n c_0^{1/n} }{ (n+1)\bar{L}} 
\, x^{\frac{n+1}{n} } } \,.
\ee
Raising both sides to the power $\frac{n}{n+1}$ and introducing
\begin{eqnarray}
H &=& \bar{H}^{\frac{n+1}{n}} \,,\\
L&= & \left[ \frac{(n+1) \bar{L}}{n c_0^{1/n} } \right]^{\frac{n}{n+1}} 
\,,
\end{eqnarray}
the Nye solution~(\ref{barredNye}) becomes the Vialov solution
\be
h(x) = H\left[ 1-\left( \frac{x}{L} \right)^{ \frac{n+1}{n}} 
\right]^{\frac{n}{2(n+1)}} \,.
\ee

In the general case in which $c(x) \neq $~const., one cannot express 
explicitly $\bar{x}$ in terms of elementary functions of $x$. If this 
happens, the reduction of the Vialov equation to the Nye equation is 
purely formal; then also the generation of the solution of the Vialov 
equation from the generating function~(\ref{barredNye}) is purely 
formal. Forms of the function $c(x)$ for which the integral $I(x)$ can be 
calculated explicitly in terms of elementary functions have been 
identified in \cite{myMG17}.

\section{Conclusions}
\label{sec:6}
\setcounter{equation}{0}

The understanding of longitudinal profiles of ice sheets, ice caps, and 
glaciers is necessary to estimate the volume of ice in these 
three-dimensional systems and the volume-area scaling relation 
\cite{FaraoniJOG16}, which is the basis to estimate the ice content of a 
region, the ice loss due to climate change, and the sea level rise due to 
the 
melting of polar ice \cite{Bahrreview}. Realistic ice follows Glen's law 
\cite{Glen55}, while its degenerate limit of perfectly plastic ice was 
used in the early days of glaciology \cite{Nye51a,Nye51b,Paterson94}. The 
longitudinal profile of an ice cap or glacier following Glen's law is 
described by the Vialov equation 
\cite{Vialov58,Paterson94,CuffeyPaterson10,Hooke05,GreveBlatter09}, in 
which the 
accumulation rate of ice due to precipitation on the glacier is described 
by the accumulation function $c(x)$. This function must be prescribed in a 
given model, ideally based on atmospheric models for the region. The 
solution of the Vialov equation is reduced to the quadrature 
of an integral $I(x)$ involving the accumulation function $c(x)$. 

It is not always possible (or easy) to express the integral $I(x)$ in 
terms of elementary functions (cf. Ref.~\cite{myMG17}) hence nothing is 
gained, in terms of solving analytically the Vialov equation, by 
introducing the new variable $\bar{x}(x)$ and rewriting the Vialov 
equation in terms of it, as we have done in Sec.~\ref{sec:2}. 
 Nevertheless, 
what is gained is a Lagrangian formulation for this equation, thus solving 
the inverse variational problem. In turn, conservation of the 
corresponding Hamiltonian leads to an analogy with a particle in 
one-dimensional motion and to an unexpected and intriguing analogy with 
the Friedmann equation of general relativistic cosmology. We have explored 
this analogy and found that, for the physically meaningful values $n>0$ of 
the Glen law exponent, the analogous universe is filled with a perfect 
fluid with ultrarelativistic equation of state $P>\rho/3$. The limit 
$n\rightarrow +\infty$ of perfectly plastic ice corresponds to a 
radiation-dominated universe with $P=\rho/3$. In addition, the 
cosmological analogy provides two new developments. First, it establishes 
that all solutions of the (rescaled) Vialov equation are 
roulettes. 
Second, it provides a new symmetry of the Vialov equation 
written in terms of the new variable $\bar{x}$. By using this idea, the 
solution of the Vialov equation for finite $n$ can be reduced, at least 
formally, to solving the Nye equation (which is trivial) and then 
expressing $\bar{x}$ in terms of $x$ (which is not). Then the Nye profile 
obtained for perfectly plastic ice \cite{Nye51a,Nye51b} becomes a 
generating function for {\em all} solutions of the Vialov equation. In 
general 
({\em i.e.}, for general forms of the function $c(x)$), this 
solution-generating technique remains purely formal because of the 
difficulty in expressing the rescaled variable $\bar{x}$ in terms of 
elementary functions of the physical variable $x$ along the glacier bed  
(but see Ref.~\cite{myMG17}).   
At least 
the Vialov profile (corresponding to $c(x)=$~const.) for Glen law ice can 
be obtained in this way from the Nye profile for perfectly plastic ice.

\section*{Acknowledgments}

We thank two referees for very helpful comments. This work is supported, 
in part, by the Natural Sciences \& Engineering Research Council of Canada 
(Grant No. 2016-03803) and by Bishop's University.

\end{document}